\documentclass[twocolumn]{aastex7}

\usepackage{physics,amsmath,amssymb,bbold,amsthm,mathrsfs}
\usepackage{xcolor,url}

\newcommand{\mki}{
  Kavli Institute for Astrophysics and Space Research, 
  Massachusetts Institute of Technology , 77 Massachusetts  Ave., 
  Cambridge, MA 02139, USA
}

\newcommand{\Change}[1]{{\color{red}\bf #1}}
\renewcommand{\Change}[1]{#1}

\begin{document}

\title{X-ray Polarimetry of Accreting White Dwarfs: A Case Study of EX Hydrae}

\author[orcid=0000-0003-2602-6703, sname=Gunderson, gname=Sean]{Sean J.\ Gunderson}
\affiliation{\mki}
\email[show]{seang97@mit.edu} 

\author[orcid=0000-0002-2381-4184, sname=Ravi, gname=Swati]{Swati Ravi}
\affiliation{\mki}
\email{swatir@mit.edu}

\author[orcid=0000-0002-6492-1293, sname=Marshall, gname=Herman]{Herman L.\ Marshall}
\affiliation{\mki}
\email{hermanm@mit.edu}

\author[orcid=0000-0003-1898-4223, sname=Swarm, gname=Dustin]{Dustin K.\ Swarm}
\affiliation{Department of Physics and Astronomy, University of Iowa, Iowa City, IA, USA}
\email{dustin-swarm@uiowa.edu}

\author[orcid=0000-0002-7204-5502, sname=Ignace, gname=Richard]{Richard Ignace} \affiliation{Department
 of Physics \& Astronomy, East Tennessee State University, Johnson City, TN 37614 USA}
\email{ignace@etsu.edu}

 \author[orcid=0000-0003-4071-9346, sname=Naze,gname=Yael]{Yael Naz\'{e}}
 \thanks{FNRS Senior research associate}
 \affiliation{Groupe d’Astrophysique des Hautes Energies, STAR, Universit\'{e} de Li\`{e}ge, Quartier Agora (B5c, Institut d’Astrophysique et de G\'{e}ophysique), All\'{e}e du 6 Ao\^{u}t 19c, B-4000 Sart Tilman, Li\`{e}ge, Belgium}
 \email{ynaze@uliege.be}
 
 \author[orcid=0000-0002-3860-6230,sname=Huenemoerder,gname=David]{David P.\ Huenemoerder}
\affiliation{\mki}
\email{dph@mit.edu}

\author[orcid=0000-0002-1131-3059,sname=Pradhan,gname=Pragati]{Pragati Pradhan}
\affiliation{Embry Riddle Aeronautical University, Department of
 Physics \& Astronomy, 3700 Willow Creek Road Prescott, AZ 86301, USA}
 \email{pradhanp@erau.edu}

\begin{abstract}
We present the first first X-ray polarization measurements of a white dwarf, the intermediate polar EX Hya. We measured significant polarization only in the 2 -- 3\,keV energy band with a polarization degree of 8 percent at a $3\sigma$ significance. No significant polarization was detected above 3\,keV, which we attribute to the higher energy bands having lower signal-to-noise. We found that the scattering surface detected by the {\em IXPE} is nearly perpendicular to the optical scattering plane, showing that the X-ray scattering surface is the WD and close to the base of the accretion column. Finally, we show how the polarization can be used to estimate the height of the accretion shock above the white dwarf's surface. 

\end{abstract}


\section{Introduction} \label{sec:Intro}

Cataclysmic variables (CVs) are close binary systems consisting of a white dwarf (WD) undergoing accretion from a main-sequence companion via Roche lobe overflow \citep{Warner1995}. \Change{When the WD is only weakly magnetic, more generally referred to as ``non-magnetic,'' the accreted matter forms an equatorial disk as it settles onto the surface of the WD. If the WD has a sufficiently strong magnetic field, the accretion disk will be disrupted in two possible ways. When the WD dwarf's spin and orbital periods are synchronized, the WD's magnetic field will accrete directly from the Roche point onto its magnetic pole(s) \citep{Schwope2025}. This type of magnetic CV (mCV) is known as a Polar. If the spin and orbital periods are asynchronous, usually with $P_\mathrm{spin} < P_\mathrm{orbit}$, an intermediate situation will occur with such objects referred to as intermediate polars (IPs). In a classical IP, an accretion disk will form, but the magnetic field disrupts it at inner radii and channels the material onto the magnetic pole(s) of the WD.}

Near the surface of the WD in an IP ($\lesssim 1\,R_\mathrm{WD}$ above the surface), the supersonic accretion material undergoes a strong shock that heats the plasma to tens of keV. The post-shock accretion column radiatively cools, making IPs profuse emitters of hard X-rays. IP X-ray spectra are especially characterized by a flat, hard thermal continuum and prominent Fe emission lines between 6--7 keV. Highly ionized Fe species in the post-shock accretion flow are evident from the presence of \ion{Fe}{25} and \ion{Fe}{26} K$\alpha$ lines at 6.7 keV and 7.0 keV, respectively \citep{Hayashi2021,Luna2018,Mukai2017}. 

Another common feature in IP X-ray spectra is a strong Fe K$\alpha$ fluorescent line at 6.4 keV \citep{Mukai2017}. The neutral Fe line is produced through a combination of reprocessed X-rays in the pre-shock accretion curtain and the surface of the WD. The primary source of the Fe K$\alpha$ emission is dependent on the height $h_0$ of the shock, but if there is a significant contribution from the WD surface, then there should be a corresponding reflection spectrum.

Some high-energy photons that fluoresce the cold Fe will be Compton-scattered and produce a unique spectral signature. \citet{Luna2018} used this unique signature to constrain an upper limit of the amount of reflected photons in the IP EX\,Hya to be $\leq15\%$ of the total spectrum. Lower energy photons will be Thomson scattered off the surface as well but are photometrically indistinguishable from the non-reflected photons. This presents a problem for constraining the total reflection amplitude using traditional X-ray observations. However, by expanding the observable parameter space to include the polarization
of the incoming photons, it becomes possible to distinguish the source and reflected photons.

Polarization is described through the polarization degree and angle
\begin{eqnarray}
    \Pi &=& \frac{\sqrt{Q^2+U^2}}{I}\\
    2\psi &=& \arctan{\frac{U}{Q}},
\end{eqnarray}
defined using the Stokes parameters $Q$ and $U$ and the total intensity $I$. One can in theory include the third stokes parameter $V$ for circular polarization, but this is not currently accessible in measurements of X-ray polarization. Polarization thus introduces two additional dimensions to our parameter space.

\citet{Matt2004} predicted that the X-rays from Polars will be polarized from both scattering within the accretion column and reflection off the WD surface. For IPs, their weaker magnetic fields and thinner accretion columns will not have significant optical depth for scattering, so there should not be appreciable polarization in source emission. The only source of polarization should therefore be from scattering off the surface, with the polarization degree acting as a direct measurement of the total reflection amplitude between both Thomson and Compton processes.

Here we report on an observation of EX\,Hya with the \textit{Imaging X-ray Polarimetry Explorer} (\textit{IXPE}) to make such a measurement. EX\,Hya presents an ideal case for this based on its orbital inclination $i_\mathrm{inc}=78\arcdeg$ and relative proximity $d=56.77$\,pc, see Table~\ref{tab:SysParams}, which together help to maximize the brightness of the source and polarization detectability in a short-duration observation. The rest of EX\,Hya's important system parameters are given in Table~\ref{tab:SysParams}, including optical polarization measurements.

\begin{deluxetable}{ccl}
    \caption{EX\,Hya system parameters \label{tab:SysParams}}
    \tablehead{
        \colhead{Parameter} & \colhead{Value} & \colhead{Reference}
    }
    \startdata
        $M_\mathrm{WD}$ ($M_\odot$) & \Change{$0.788\pm0.025$} & (a)\\
        $M_\mathrm{M}$ ($M_\odot$)& \Change{$0.1074\pm0.0047$} & (a)\\
        $P_\mathrm{spin}$ (s) & $4021.6$ & (a) \\
        $P_\mathrm{orbit}$ (s) & $5895.4$ & (a) \\
        $i_\mathrm{inc}$ (deg) & \Change{$78.0\pm0.2$} & (a) \\
        $d$ (pc) & $56.77\pm0.05$ & (b) \\
        $\Pi_\mathrm{optical}$ (\%) & \Change{$0.3\pm0.1$} & (c) \\
        $\psi_\mathrm{optical}$ (deg) & $71\pm13$ & (c) \\
    \enddata
    \tablerefs{(a) \citet[][and citations therein]{Beuermann2003}, (b) \citet{BailerJones2021}, (c) \citet{Cropper1986}}
\end{deluxetable}

In \S~\ref{sec:DataRed}, we provide details on our data reduction pipeline and modeling methods. In \S~\ref{sec:PolMeasure}, we discuss our measurements of the polarization signal and potential variabilities. In \S~\ref{sec:ShockHeight}, we estimate the height of the shock in EX\,Hya's accretion column from the polarization measurement. Finally, in \S~\ref{sec:Conclusions}, we give our conclusions.

\section{Data Reduction} \label{sec:DataRed}
\textit{IXPE} is an X-ray observatory launched in 2021 by NASA and ASI to perform imaging X-ray polarimetry in the 2--8 keV band \citep{2022JATIS...8b6002W}. \textit{IXPE} is made up of three identical telescopes consisting of Wolter-I mirror assemblies and gas pixel detectors \citep{2021AJ....162..208S, 2021APh...13302628B}.

\textit{IXPE} observed EX Hya between 2025 January 18-26, with a 286.4\,ks exposure for each of the three detector units under the observation ID 04003999. \textit{IXPE} data was processed with the standard pipeline in \textsc{heasoft} v6.34. We additionally made use of community-provided software\footnote{\url{https://heasarc.gsfc.nasa.gov/docs/ixpe/analysis/contributed.html}} for background event rejection \citep{DiMarco2023} and general filtering and binning using \textsc{ixpeobssim} \citep{Baldini2022}.

\section{Polarization Measurements}\label{sec:PolMeasure}

We applied two methods for measuring a polarized signal: a model-independent polarimetric analysis and event-based Bayesian nested sampling. We extracted a $60\arcsec$ radius region centered around the source at $\mathrm{RA} = \mathrm{12:52:25.6134}$, $\mathrm{Dec}=\mathrm{-29:14:53.081}$, obtaining a source count rate of 0.08 $\mathrm{counts\;s^{-1}\;arcmin^{-2}}$ after both background rejection and subtraction per the prescription of \cite{DiMarco2023}. Background subtraction was estimated utilizing an annulus of inner radius 150" and outer radius 300" centered at the source. We used \texttt{ixpeobssim} software version 31.0.3 for model-independent polarimetric analysis using the weighted polarization cube (PCUBE) algorithm \citep{KISLAT201545,Baldini2022}. 

The second method, utilizing an identical background treatment, was an event-based maximum likelihood analysis described in \citet{Herman2021a,Herman2021b,Herman2024} and implemented as a Bayesian nested sampling (BNS) method in the \Change{{\sc queen-bee} software by \citet{ravi2025queenbee,Ravi2025}}.\footnote{Software is available via GitHub: \url{https://github.com/swati-ravi/QUEEN-BEE}} The BNS method characterizes each model using a log-evidence parameter $\ln Z$, where $Z$ is the likelihood. Model comparison between two hypothetical Models 1 and 2 using BNS involves the standard Bayesian statistical practice of calculating the difference $\Delta\ln Z_{12} = \ln Z_1 - \ln Z_2$ and then computing a Bayes factor defined as 
\begin{equation}
    B_{12}=\exp(\Delta\ln Z_{12}).
\end{equation}
The significance of this Bayes factor is then evaluated using Jeffreys' Scale \citep{Jeffreys1986}. When using this BNS method, we compared a constant polarization model against an unpolarized model.

The results of these two methods are summarized in Table~\ref{tab:PolMeasures}. We also provide the MDP$_{99}$, the minimum detectable polarization that has only a 1 percent chance of being a false alarm. In the standard analysis of X-ray polarimetry, a polarization signal is only considered to be detected if the measured polarization degree is above MDP$_{99}$.

\begin{deluxetable*}{cc|ccc|ccc}
    \tablecaption{Polarization measurements with 68 percent uncertainties after background rejection and subtraction in different energy bands.}\label{tab:PolMeasures}
    \tablehead{
        \multicolumn{2}{c}{}& \multicolumn{3}{c}{PCUBE} & \multicolumn{3}{c}{Bayesian Nested Sampling}\\
        \colhead{$E$ (keV)} & \colhead{MDP$_{99}$ (\%)\;\tablenotemark{a}} & \colhead{$\Pi$ (\%)} & \colhead{$\psi_\mathrm{X-ray}$ (deg)} & \colhead{$\sigma$\;\tablenotemark{b}} & \colhead{$\Pi$} & \colhead{$\psi_\mathrm{X-ray}$} & \colhead{$B_\mathrm{pol}$\,\tablenotemark{c}}
    }
    \startdata
    2 -- 8 & 5.2 & $4.4\pm1.8$ & $-38.9\pm11.5$ & 2.20 & $3.6_{-1.3}^{+1.6}$ & $40.8_{-12.2}^{+10.1}$ & 0\;\tablenotemark{d} \\
    \textbf{2 -- 3} & \textbf{6.4} & $\mathbf{8.0\pm2.2}$ & $\mathbf{-34.2\pm7.7}$ & \textbf{3.25} & $\mathbf{6.9_{-2.7}^{+2.8}}$ & $\mathbf{-42.8_{-10.5}^{+10.0}}$ & $\mathbf{12.8\pm3.8}$ \\
    3 -- 4 & 7.0 & $4.2\pm2.3$ & $-44.1\pm16.1$ & 1.07 & $4.8_{-2.6}^{+3.8}$ & $38.8_{-19.3}^{+17.3}$ & $0.03\pm0.01$ \\
    4 -- 6 & 8.6 & $2.4\pm2.9$ & $70.3\pm34.5$ & 0.26 & $3.9_{-2.6}^{+4.8}$ & $-11.7_{-21.3}^{+26.5}$ & $0.004\pm0.001$ \\
    6 -- 8 & 19.8 & $4.9\pm7.1$ & $-39.8\pm41.1$ & 0.50 & $5.5_{-2.9}^{+8.6}$ & $35.5_{-32.2}^{+30.0}$ & $0.008\pm0.002$ \\
    \enddata
    \tablenotetext{a}{Polarization degree that that has a 1 percent chance of being a statistical fluke.}
    \tablenotetext{b}{Statistical significance of the PCUBE polarization signal.}
    \tablenotetext{c}{Bayes factor comparing the constant polarization model to no polarization.}
    \tablenotetext{d}{Formally, the Bayes factor for the 2 -- 8\,keV band is $B_\mathrm{pol}=\mathrm{e}^{-11000}$. This number is too small to give precise errors.}
\end{deluxetable*}

We first looked at the polarization signal across the entire 2 -- 8\,keV band pass of \textit{IXPE}, which has a minimum detectable limit of MDP$_{99} = 5.2\,\%$. The PCUBE measurement does not cross this threshold, so we conclude that in this energy band there is no detectable polarization. The BNS measurement reinforces this as the Bayes factor is decisively in preference for the unpolarized model.

We next applied the PCUBE and BNS methods in narrower energy bands to search for energy-dependence in the polarization signal. The softest energy band we used, 2 -- 3\,keV, showed a substantial polarization degree of $\Pi=8.0\pm2.2\,\%$ at a $3\sigma$ significance for PCUBE and a decisive measurement of polarization, based on the Bayes factor of $B_\mathrm{pol}>10$, with a polarization degree of $\Pi=6.9_{-2.7}^{+2.8}\,\%$ for BNS. Both values appear above the MDP$_{99}$ threshold. The other energy bands, 3 -- 4, 4 -- 6, and 6 -- 8\,keV, do not show statistically significant levels of polarization.


The source of this energy dependence is interesting since we expect the polarization to come from Thomson scattering off the surface of the WD, which should be an energy-independent process. We propose that our energy dependence is instead due to the distribution of radiation from the accretion column. In the standard picture of the post-shock accretion column \citep[see for example][]{Hayashi2014,Hayashi2018}, the gas cools from a maximum temperature at the top of the column to some lower bound as it settles on the WD surface. This distribution of temperature with height $T(h)$, where $0 < h < h_0$ as measured from the surface of the WD, results in different portions of the column emitting a different distribution of energies at each height. In other words, the column can be partitioned into a distribution of X-ray sources, each giving a unique spectrum $S_E(T,h)$.

Following the description given in \citet{Hayashi2014}, we will consider the column being partitioned into a number of equilibrium thermal plasma sources. Near the top of the column $T(h_0)\approx T_\mathrm{sh}$, so high energy photons will be emitted mostly near $h_0$.\footnote{In reality, cooler portions of the post-shock accretion column can still emit higher energy photons, but the relative number of photons will be substantially smaller due to the exponential cutoff of thermal emission above the plasma temperature. This allows us to say to first order that only the top of the column gives us high energy photons.} Further down the column, the temperature drops, shifting the peak of the emission measure to lower and lower energies. At all heights, however, lower energy photons (i.e., 2 -- 3\,keV) are emitted, increasing the total number of scattering photons and improving the signal in this band. The non-detection at higher energies may thus not come from a lower polarization degree at high energies but from the lower signal-to-noise, as reflected in the increased MDP$_{99}$ values with energy. A detailed spectropolometric analysis will need to be conducted to rigorously determine whether the polarization has any real energy dependence.

\begin{figure}
    \centering
    \includegraphics[width=\linewidth, trim=0 10 0 10]{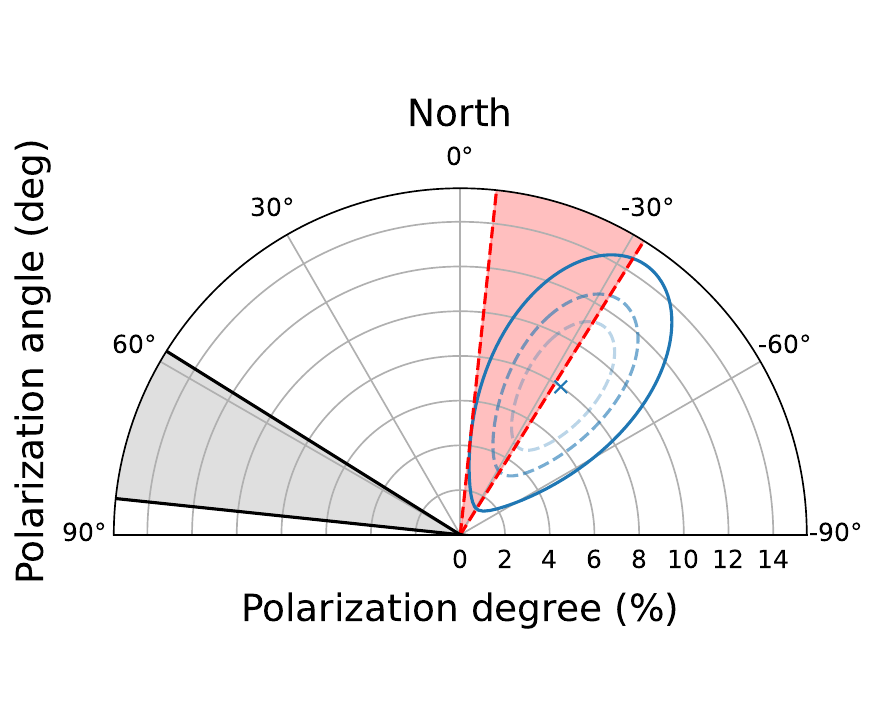}
    \caption{\Change{Protractor plot of our X-ray polarization measurement (specifically PCUBE; blue contours) in the 2 -- 3\,keV band. Contours are the 68, 90, and 99 percent confidence contours. Grey shaded region is the confidence in the optical polarization. The red shaded region is the direction of the magnetic field pole based on the optical polarization.}}
    \label{fig:ProtractorPlot}
\end{figure}

The next quantity of interest is the polarization angle in the 2 -- 3\,keV band. We show the relative position angle of the X-ray polarization on the sky in Figure~\ref{fig:ProtractorPlot} in blue. As a point of reference, we also show the optical polarization angle given in Table~\ref{tab:SysParams} in orange, which shows the direction of the cyclotron scattering plane \citep{Cropper1986} and thus the direction of the magnetic poles on the sky (red dashed lines).

\Change{The optical polarization measurement by \citet{Cropper1986} was taken over a 200\,min exposure, so it shows the average position of the magnetic field over both the spin and orbital periods. The same is true for the X-ray polarization angle since the \textit{IXPE} observation was taken over a total of 8\,days. Thus the discussion that follows concerns the spin- and orbital-average behavior of the accretion geometry.}

Based on the angles in Figure~\ref{fig:ProtractorPlot}, the X-ray scattering plane is nearly perpendicular to the magnetic field. Two details can be further gleaned from this. First, the X-ray scattering surface is not within the accretion column but off the WD surface as predicted. If the polarization were due to scattering within the column, we would have measured $\psi_\mathrm{X-ray}$ to be parallel to $\psi_\mathrm{Optical}$. More importantly, the second detail is that the scattering surface is relatively close to the accretion column and not at a far latitude from the magnetic pole.

\begin{figure*}
    \centering
    \includegraphics[width=\linewidth,trim=0 85 0 90]{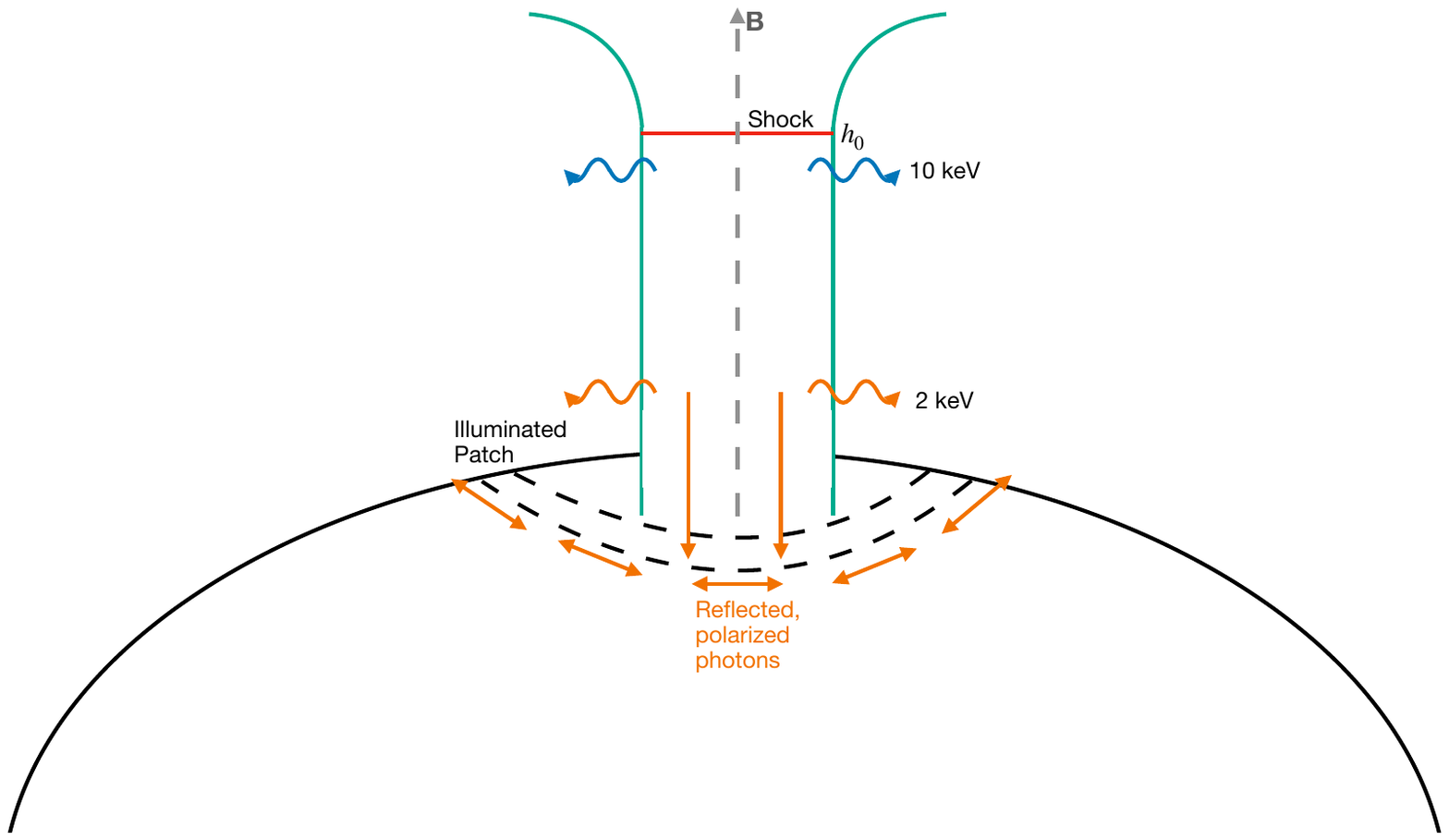}
    \caption{Diagram of the accretion column and scattering geometry in EX Hya. The observer is out of the page, looking at the accretion column perpendicularly. On the sky, this would be orientated 34.2 degrees to the northeast.}
    \label{fig:AccretionDiagram}
\end{figure*}

If the illuminated surface were far from the accretion column's base, we would find a much larger difference in the polarization angle from the magnetic field's position in Figure~\ref{fig:ProtractorPlot}. There is some difference, given that scattering is unlikely to be at the exact location of the magnetic polar cap, but it may also suggest a not-insignificant width to the illuminated ring around the accretion column. A simple schematic diagram of our inferred scattering geometry is shown in Figure~\ref{fig:AccretionDiagram}. The illuminated patch, or rather ring in our assumption, is marked by the black dashed lines on the surface of the WD. What we measure for the polarization is the average direction of the polarized, reflected photons, which come out to the same angle as those directly in front of the column.

\Change{It is worth noting the simplifications that we have used in making Figure~\ref{fig:AccretionDiagram}. The accretion geometry in EX\,Hya is better described as being a ``curtain'' that stretches over many longitudes \citep{Echevarria2016,Bisikalo2020}. The asymmetry of this curtain should produce a variable polarization signal based on the (vertical) cross-sectional shape we view at any given spin and orbital phase. A spin/orbital periodic polarization degree was also predicted by \cite{Matt2004} for the reflection component of the polarization in Polars due to their spin, so we next look for possible variability in the X-ray polarization.}

\begin{figure}
    \centering
    \includegraphics[width=\linewidth]{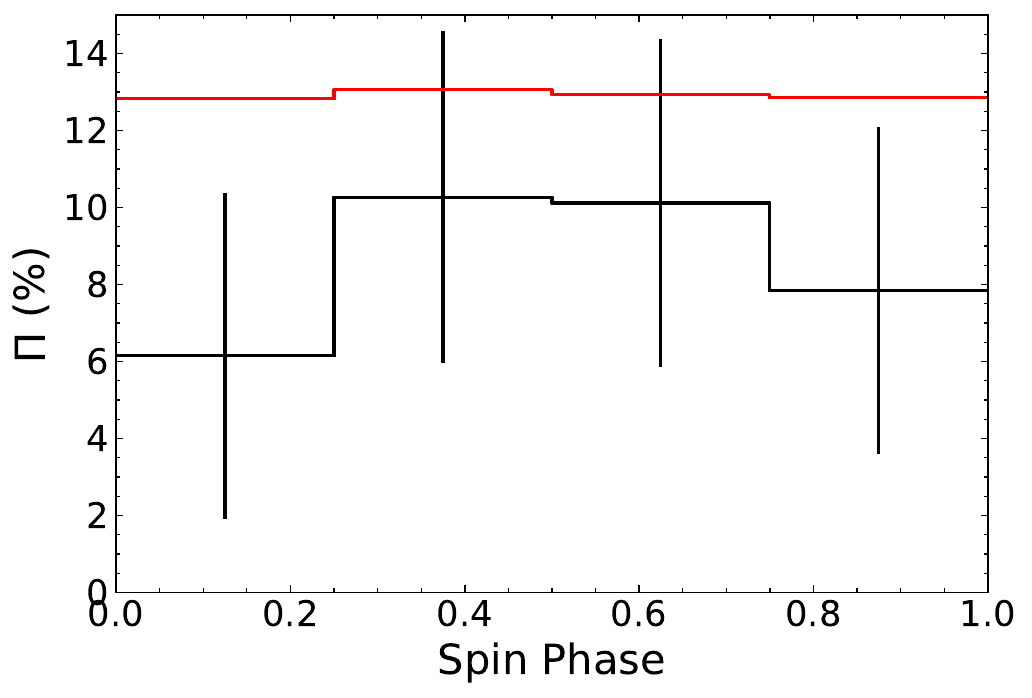}
    \caption{Polarization degree $\Pi$ in 0.25 phase bins for the 2 -- 3\,keV band against spin phase with 68 percent uncertainties (black histogram). MDP$_{99}$ values for each bin are plotted in red.}
    \label{fig:PhaseResolvedPD}
\end{figure}

To do so, we used \textsc{ixpeobssim} to phase-fold the data over EX\,Hya's spin period. We note, though, that EX\,Hya's spin period is measurably decreasing, so we used the quadratic ephemeris from \citet{Beuermann2024}
\begin{equation}
    t_\mathrm{max} = 2437699.89333 + 0.0465464497C - 5.61\times10^{-13} C^2,\label{eq:spinup}
\end{equation}
where $C$ is the spin cycle and times are given in BJD. The resulting spin-phase resolved polarization measurements for the 2 -- 3\,keV band are shown in Figure~\ref{fig:PhaseResolvedPD}, where we used bins of a quarter phase.

There does not appear to be any concrete evidence of variability in the X-ray polarization (black histogram) due to each bin being below its associated MDP$_{99}$ (red). As such, none can be considered a significant polarization detection. \Change{Consequently, we can not consider the more detailed geometry of the accretion curtain. We are limited to only consider the spin- and orbit-averaged cross-sectional shape that the curtain will have based on our line of sight. This will be that of a cylindrical column as we show in Figure~\ref{fig:AccretionDiagram}.}

\Change{We also tested for linear polarization rotation, i.e., where the polarization degree is constant but its angle rotates in the $Q-U$ plane. \textsc{queen-bee} includes a model for testing linear polarization rotation with time \citep{Ravi2025}. The results for the 2--3~keV bin show weak preference for polarization rotation of $-54_{-5}^{+4}$\,deg\,day$^{-1}$ with a Bayes Factor of $B=2.38_{-3.84}^{+1.48}$ compared with the constant polarization model. This is only a weak detection, though, so we can not confidently say the rotation is detected. A longer observation of EX\,Hya, either with \textit{IXPE} or a future soft X-ray polarization mission will be needed to determine any polarization variability in degree and/or angle.}

\section{Estimating Shock Height}\label{sec:ShockHeight}

As noted in \S~\ref{sec:Intro}, the height of the shock $h_0$ is the fundamental quantity that drives the strength of the Fe K$\alpha$ line. This makes the Fe K$\alpha$ line a useful diagnostic for determining the shock height \citep[e.g.,][]{Hailey2016,Mukai2017}, but the dual source of Fe K$\alpha$ emission complicates the diagnostic. Reprocessed photons from the column and the surface are not distinguishable. More complex models, such as the comprehensive IP accretion emission and reflection models \citep{Hayashi2014,Hayashi2018}, can estimate shock heights hydrodynamically \citep{Hayashi2021}, but there remain degeneracies, since different shock geometries can produce the same spectral shape. Time-dependent spectral analyses have been proposed to break this degeneracy \citep{Belloni2021}, but this method requires detailed 3D modeling and its feasibility has not been demonstrated.

A more direct approach is to use the reflection amplitude since that will directly depend on the height without confusion with other sources. The previously cited work by \citet{Luna2018} determined the amplitude of the Compton reflected photons to be up to 15 percent through a detailed spectral analysis of available \textit{NuSTAR}, \textit{Swift}, and \textit{Chandra}/HETG data. They then concluded, from just Compton scattered photons, that that the shock height in EX\,Hya is $h_0/R_\mathrm{WD}\leq0.9$. At the same time, their thermal shock constraints predicts a narrower range $0.3\lesssim h_0/R_\mathrm{WD}\lesssim 0.75$. The range is based on assumptions of the inner radius of EX\,Hya's accretion disk. 

The question we ask now is whether $h_0$ can be discerned from the polarization signal to further refine the estimates from \citet{Luna2018}. The geometry of the emitting column and reflection is similar to that of the ``lamp post'' model for black holes \citep{Reynolds2014}, i.e., an emitting source raised a distance $h_0$ above the reflecting surface. Given this similarity, we can adapt lamp post models, such as the polarization model from \citet{Dovciak2004}.

As noted before, the observed reflected photons observed by \textit{IXPE} are Thomson scattered, so the reflected spectrum will be the same shape as the direct emission but a smaller normalization $S_\mathrm{Ts}(E) = \eta N_\mathrm{i}S(E)$, where $\eta=S_\mathrm{Ts}/S_\mathrm{i}$ is the ratio of the amount of flux scattered into our line of sight from the incidence light on the reflecting surface and $N_\mathrm{i}$ is the proportion of light incident on the scattering surface from the source. Following \citet{Dovciak2004}, this proportion of incident light is
\begin{equation}
    N_\mathrm{i} = N_\mathrm{source} \frac{R_\mathrm{WD}^2}{r}\sin\theta_h \dv{\theta_h}{r}.\label{eq:NiStart}
\end{equation}
Note that we have corrected for the general relativistic terms that would not apply to the case of a white dwarf. Here $\theta_h$ is the angle of incident emission relative to the accretion column, and $r$ is the latitudinal radius where the reflection occurs.

Assuming a simple 2D geometry that does not account for the curvature of the WD, the radius of reflection is related to the angle of emission by $r=h\tan\theta_h$. This allows us to simplify Equation~\ref{eq:NiStart} to
\begin{equation}
    \frac{N_\mathrm{Ts}}{\eta N_\mathrm{source}} = \frac{R_\mathrm{WD}^2h}{\left(h^2+r^2\right)^{3/2}},\label{eq:NTsEnd}
\end{equation}
where we have also used our earlier statement that $N_\mathrm{Ts}/N_\mathrm{i} = \eta$. By definition, the polarization is the ratio of polarized, i.e. reflected, light to the total observed light, so the polarization fraction will be $\Pi = N_\mathrm{Ts}/\eta N_\mathrm{source}.$ In other words, Equation~\eqref{eq:NTsEnd} is the polarization fraction of the reflected photons from a particular point $h$ in the accretion column.

If we break the column into $K$ number of isothermal point sources, then each point source $i$ will have their own polarization $\Pi_\mathrm{i} = \sqrt{Q_\mathrm{i}^2+U_\mathrm{i}^2}/I_\mathrm{i}$. The polarization we measure will then be the average from the entire column. One difficulty for this calculation is that polarization values can not be simply averaged due to the their vector definition.

The symmetry in our system, however, allows us to overcome this problem. From the perspective of the accretion column shown in Figure~\ref{fig:AccretionDiagram}, the polarization will be 
\begin{equation}    
\Pi_{\mathrm{i},*} = \frac{\sqrt{Q_{\mathrm{i},*}^2 + U_{\mathrm{i},*}^2}}{I_{\mathrm{i}}} = \frac{\left|Q_{\mathrm{i},*}\right|}{I_\mathrm{i}}.
\end{equation}
Note though that just as the intensity $I_\mathrm{i}$ is independent of the orientation on the sky, so to is the overall polarization $\Pi_\mathrm{i} = \Pi_{\mathrm{i},*}$. Furthermore, because we found that the scattering angle is perpendicular to the accretion column, in this column-orientated perspective, $|Q_{\mathrm{i},*}| = Q_{\mathrm{i},*}$ for all point sources.

We can then say that from the perspective of the accretion that the average polarization will then be
\begin{equation}
    \Bar{\Pi} = \sum_{\mathrm{i}=1}^K w_\mathrm{i}\frac{Q_{\mathrm{i},*}}{I_\mathrm{i}} = \sum_{\mathrm{i}=1}^K w_\mathrm{i} \Pi_\mathrm{i},
\end{equation}
where $w_\mathrm{i}=I_\mathrm{i}/I$ is the intensity-weighting of the $i$-th source. The result of the sum $\Bar{\Pi}$ is the value we measure in the \textit{IXPE} data from our perspective. It is only due to our knowledge of the orientation of the scattering planes between optical and X-ray components that we can decompose the system into just the $Q_{\mathrm{*},i}$ parameter.

If we let $K\rightarrow\infty$, then the sum will converge to an integral over the height of the column
\begin{equation}
    \Bar{\Pi} = \int_0^1 w(a)\frac{\mathrm{d}\Pi}{\mathrm{d}a}\mathrm{d}a.\label{eq:meanPiInt}
\end{equation}
Here we have converted to a unitless scale parameter $a\equiv h/h_0$ and $\Pi$ now takes the form of Equation~\eqref{eq:NTsEnd}. In this form, we are assuming that the scattering surface can be characterized to occur at a single latitudinal radius on the WD, i.e., it does not have a large extent. Based on the X-ray and optical polarization angles, the center of the scattering surface will be $15.2\pm7.7$\,degrees from the accretion column. If the accretion column and the geometric pole of the WD do not have a large offset, this will correspond to a radius of $r=0.26\pm0.13\,R_\mathrm{WD}$.

The weights of the integral will functionally be the same but now depend of height
\begin{equation}
    w(a) = \frac{I_{\Delta E}(a)}{I_{\Delta E,\mathrm{Total}}} = \frac{L_{\Delta E}(a)}{L_{\Delta E,\mathrm{Total}}}\label{eq:weightsDefinition}
\end{equation}
where the values are specifically over the energy range $\Delta E$ that polarization has been measured. In our case that will be 2--3\,keV. By definition, the luminosity is
\begin{equation}
    L_{\Delta E}(T) = \int_{\Delta E} \epsilon_E(T) \mathrm{DEM}(T)\mathrm{d}E,
\end{equation}
where $\epsilon_E(T)$ is the emissivity of the plasma and $\mathrm{DEM}$ is the differential emission measure. While simple as a function of temperature, we need the dependence of $L_{\Delta E}$ on $a$ for our computation.

In the isobaric assumption of the accretion flow, we can directly connect the height in the column to each temperature using
\begin{equation}
    a = \frac{\int_0^T T^{'2}/\Lambda(T')\mathrm{d}T}{\int_0^{T_\mathrm{sh}} T^{'2}/\Lambda(T')\mathrm{d}T}.\label{eq:aofT}
\end{equation}
This assumes that $\mathrm{d}T/\mathrm{d}t = n_\mathrm{sh}T_\mathrm{sh}^2\Lambda(T)/5k_BT^2$, where $n_\mathrm{sh}$ and $T_\mathrm{sh}$ are the density and temperature right after the shock and $\Lambda(T)$ is the cooling function. See \citet{Luna2015} for further details on this hydrodynamics of an isobaric flow.

The emission measure from an isobaric flow depends on the radius of the accretion column $R_\mathrm{col}$, but if we assume a simple cylindrical geometry it can be written as
\begin{equation}
    \mathrm{DEM}(T) = \frac{5}{4}\pi R_\mathrm{col}^2 v_\mathrm{ff}\frac{k_B}{\Lambda(T)},
\end{equation}
where $v_\mathrm{ff}$ is the free fall velocity of the WD \citep{Luna2015}. This allows us to give a full expression for the weights from Equation~\eqref{eq:weightsDefinition} as
\begin{equation}
    w(a(T)) = \frac{\int_{\Delta E} \epsilon(T')/\Lambda(T')\mathrm{d}E}{\int_{\Delta E}\int_0^{T_\mathrm{sh}} \epsilon(T')/\Lambda(T')\mathrm{d}T\mathrm{d}E}.\label{eq:WeightsFull}
\end{equation}
Thus to solve for the height of the shock $h_0$, the total system of equations is Equations~\eqref{eq:meanPiInt}, \eqref{eq:aofT}, and \eqref{eq:WeightsFull}.

To solve these equations, one needs to know both the shock temperature and plasma characteristics, like elemental abundances. This can be achieved with detailed spectral fitting, but for our purposes we will use the shock temperature $k_\mathrm{B}T = 19.7$\,keV and $\mathrm{Fe}/\mathrm{F_\odot}=0.88$ abundance from \citet{Luna2018} to show the efficacy of this method.

For the emissivity and cooling function, we made use of the APED \citep{Foster2012,Smith2001} accessed through the Interactive Spectral Interpretation System \citep[\textsc{isis;}][]{Houck2000} Using our measured polarization from PCUBE of $8.0\pm2.2$, the shock height above surface in EX\,Hya is $h_0=0.53\pm0.1\,R_\mathrm{WD}$.

This shock height is fully consistent with the analysis from \citet{Luna2018}, showing the efficacy of using X-ray polarization for this purpose. Importantly, though, our measurement is independent of any assumptions of the inner radius of the accretion. Our calculated shock height, in as much as our simplifying assumption can be be trusted, can be considered a more precise measure. Furthermore, we can also use our shock height as an independent parameter to say, based on \citet{Luna2018}, that the inner radius of the accretion disk is approximately $10\,R_\mathrm{WD}$.

\section{Conclusions} \label{sec:Conclusions}

We presented the results of the first accreting white dwarf observation with \textit{IXPE}, focusing on the IP EX\,Hya. Our analysis found a significant polarization signal of 6 -- 10\% from only 2 -- 3\,keV photons. For energies between 3 -- 8\,keV, no statistically significant polarization is detected due to the low signal-to-noise. The highest energy photons are, characteristically, only emitted near the top of the column where the hottest plasma occurs. In contrast, the softer photons are emitted from the entire column and are thus more numerous.

From the detected polarization signal, we determined the relative geometry of the X-ray scattering plane to the optical scattering plane. These two planes are nearly perpendicular, showing that the scattering is not occurring in the column but instead on the WD surface. Moreover, the small offset between the magnetic field and X-ray polarization angles suggest that the scattering surface on the WD is not far from the accretion column. 

We then explored possible variability in the polarization due to the WD's spin. We found that the polarization degree measurements were below $\mathrm{MDP}_{99}$ for each phase. As such, no spin variability could be determined from our data. Future observations should be planned to increase the signal-to-noise such that variabilities can be investigated.

Finally, we estimated the shock height of the accretion column from our measured polarization in the 2--3\,keV band. We used a modified black hole lamp post model for this and found a shock height of $h_0=0.53\pm0.1\,R_\mathrm{WD}$. This is consistent with the recent determinations from \citet{Luna2018} but has no dependence on other parameters like the accretion disk's truncation radius.

Our measurements highlight the need for future work on EX\,Hya and polarization from accreting WDs in general. For EX\,Hya, the analysis shown here should be expanded into a full spectropolarimetric analysis. Combining our \textit{IXPE} data with archival data used by \citet{Luna2018} will allow for an even more precise determination of the polarization across the entire 2--8\,keV band. This will allow a more concrete determination of the shock height. Additionally, our system of equations for solving for $h_0$ could be improved with more detail. For example, the reflecting surface, i.e. the WD surface, is curved and over a extended area. A full 3-dimensional computation will be needed to properly compute how a ray scatters based on both its emitting height and the location on the surface it scatters at.

In terms of observations, EX Hya should be reobserved by \textit{IXPE} for a longer total exposure time to better constrain the possible variability in the polarization signal. Based on Figure~\ref{fig:PhaseResolvedPD}, a minimum of $\sim$150\,ks exposure time is needed to measure $\Pi$ in spin phases 0.25 -- 0.75, corresponding to 600\,ks total exposure. For the other two bins, a total exposure time of $\sim$1.2\,Ms is needed, which is within nominal \textit{IXPE} observation times for other objects.

The strong soft X-ray polarization shows that accreting white dwarfs are potential high quality targets for soft X-ray polarimeters, such as the upcoming \textit{REDSoX} sounding rocket mission \citep{Herman2017} and/or the proposed \textit{GOSoX} mission \citep{Herman2021GOSoX}. Other classes of accreting WDs (e.g., non-magnetic CVs, Polars, and Symbiotic Stars) should be observed with \Change{X-ray polarimetric missions to determine their accretion geometries with high precision}. Our analysis is the first to show the diagnostic potential X-ray polarimetry has to answer such questions about accreting WD systems.

\begin{acknowledgments}
Support for SJG, HLM, and DPH was provided by NASA through the Smithsonian Astrophysical Observatory (SAO) contract SV3-73016 to MIT for Support of the Chandra X-Ray Center (CXC) and Science Instruments. CXC is operated by SAO for and on behalf of NASA under contract NAS8-03060.

SJG also received support from NASA through IXPE award number 80NSSC25K0093 issued by NASA.

YN acknowledges FNRS for support.

\end{acknowledgments}

\facilities{IXPE}

\software{ixpeobssim \citep{KISLAT201545,Baldini2022}, \textsc{isis} \citep{Houck2000}, \textsc{queen-bee} \citep{ravi2025queenbee}
          }

\bibliography{bib}{}
\bibliographystyle{aasjournalv7}

\end{document}